# Application of Doppler Broadened Gamma Spectroscopy to Study the Surface of Graphene

A. H. Weiss[1, a)], V. A. Chirayath[1, b)], R. W. Gladen[1], A. J. Fairchild[1], M. D. Chrysler[1], P. A. Sterne[2], and A. R. Koymen[1]

[1]*Department of Physics, University of Texas at Arlington, Arlington, Texas – 76019*
[2]*Lawrence Livermore National Laboratory, Livermore, California 94550*

[a)]*Corresponding author: weiss@uta.edu*
[b)]*chirayat@uta.edu*

**Abstract.** We present Doppler broadened gamma spectra, obtained using the newly developed advanced positron beam at the University of Texas at Arlington, from a sample consisting of 6 to 8 layers of graphene (MLG) on polycrystalline Cu. The kinetic energy of the positron beam was varied form 2 eV to 20 keV allowing for a depth resolved measurement. The ratio curves formed by dividing the measured Doppler broadened gamma spectra obtained at low positron kinetic energies (~2eV) to the gamma spectra obtained at 20 keV were compared to ratio curves found by dividing the calculated spectra of bulk graphite to bulk Cu. The ratio curves obtained from the measured results show qualitative agreement with those obtained from the calculated spectra. In particular, both sets of curves indicate a much reduced intensity at high momentum. The agreement between the measured and calculated curves is consistent with the hypothesis that the 2eV spectra correspond to the Doppler broadened spectra from the thin overlayer of Graphene (which we anticipate should be similar to the spectra obtained from bulk graphite) and that the spectra taken at 20 keV corresponds to bulk Cu due to the fact that most of the positrons implanted at this energy annihilate in the Cu substrate. The results taken at 2 eV provide evidence that it is possible to obtain chemically sensitive information from the top atomic layers of surfaces (both internal and external) from an analysis of the high momentum tail of the Doppler broadened gamma spectra obtained from the annihilation of positrons at the surface.

## INTRODUCTION

Doppler Broadened Gamma Spectroscopy has found wide applications in bulk studies of defects [1]. Previous theoretical and experimental work by Alatalo et al. [2], Asoka-Kumar et al. [3] and Brusa et al. [4] have shown that it is possible to use Doppler broadening of the annihilation gamma to determine the local chemical environment of the positron at the time of the annihilation in bulk samples. The selective probing of surfaces by positrons and the fact that the annihilation gamma carry the positron-electron momentum information has been utilized to investigate the surface composition and electronic structure of embedded quantum dots (QD) and nano-crystals (NC) [5, 6]. For example, Chai et al. [6] utilized beam based 2d-Angular correlation of annihilation radiation (ACAR) to study the surfaces of PbSe NC embedded inside multi layers and found that the positrons annihilated mainly with the Se electrons at the NC surfaces.

In this paper, we report initial experiments aimed at exploring the utility of Doppler broadened gamma spectroscopy to study the chemical nature of the top most atomic layer of the sample by using very low energy (~1eV - 10 eV) positrons. Previous experiments including those in our laboratory employing positron annihilation induced Auger spectroscopy indicate that almost all of the positrons implanted at these low energies either become trapped in the surface state or form positronium. The positrons that trap in the surface state annihilate with electrons, (including core electrons) of atoms in the topmost atomic layer. This fact suggests that it should be possible to use the information in the high energy tail of the Doppler broadened annihilation gamma rays to obtain information about the surface chemical composition as it has only a little contribution from the annihilation spectrum of a Ps atom (mainly para Ps). Utilizing the high energy region of the annihilation gamma from surface trapped positrons we were able to identify a few layers of carbon on polycrystalline Cu substrate.

## EXPERIMENTS

The experiments were carried out using the new long flight path (~ 3 m) TOF PAES system at the University of Texas at Arlington [7]. A brief description of the advanced positron beam system is given below and the schematic is shown in Fig.1. The variable energy positron beam uses a high efficiency rare gas moderator system. The beam energy can be tuned from 2 eV to 20 keV by appropriately biasing the moderator and the sample. The advanced beam system is equipped to measure the energy of the positron annihilation induced Auger electrons and the Doppler broadening of the 511 keV produced as a result of the annihilation in coincidence. The positron and electron paths are separated using an E X B field. The annihilation gamma spectra were measured using a 2 inch high purity Ge detector (HpGe) (Ortec GEM10195) that had an energy resolution of ~ 0.98 keV at 356 keV of Ba-133. The HpGe spectra was collected in coincidence with a 2" NaI (Tl) detector. The sample consisted of multiple layers of graphene (~ 6 – 8 layers) grown on a polycrystalline Cu foil (MLG) purchased from ACS materials. The presence of MLG on Cu was confirmed using Raman spectroscopy and PAES [8].

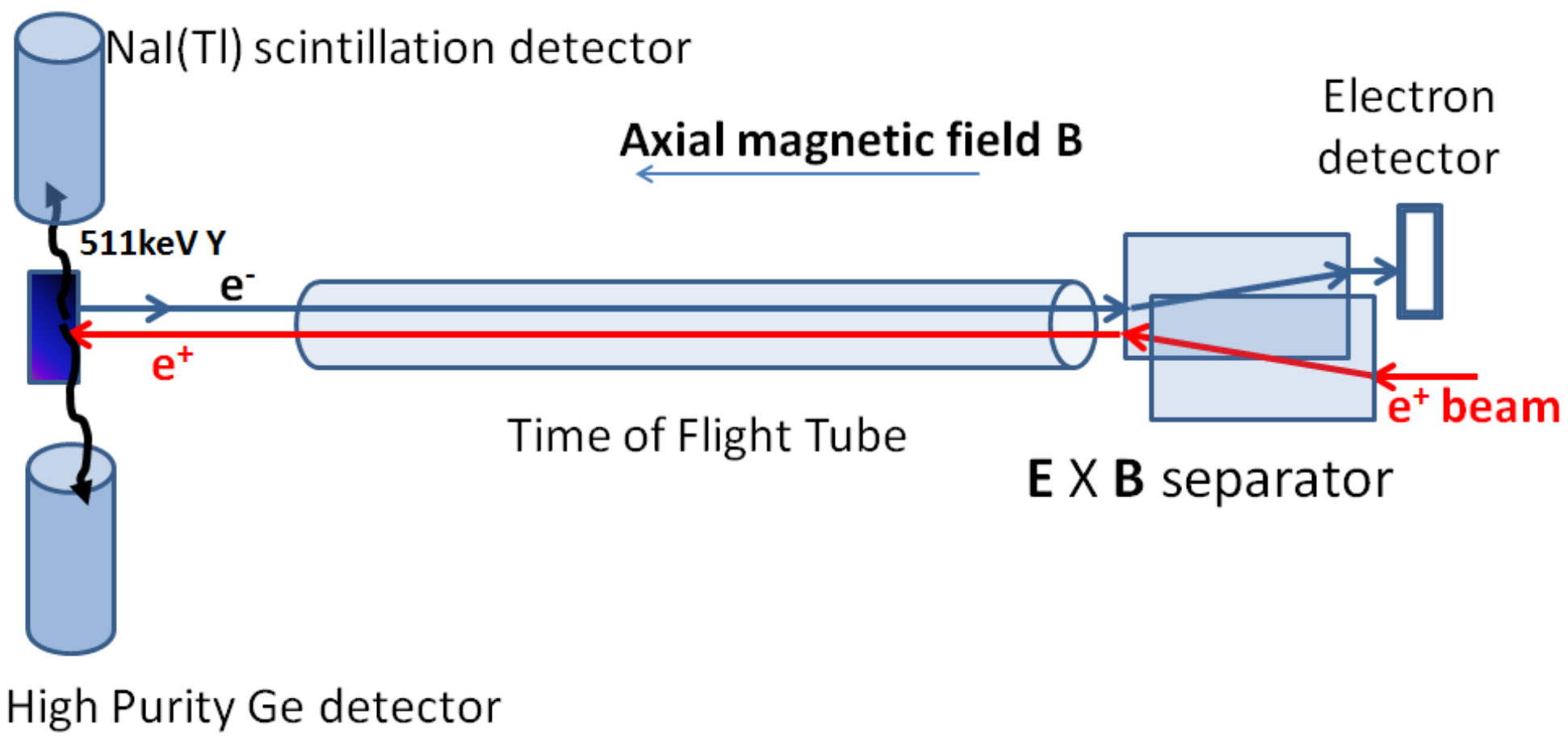

**FIGURE 1.** Apparatus used in obtaining the Doppler Broadened Gamma Spectra from the surface of Cu with multiple (6-8) Graphene over-layers. The advanced positron beam has a ~ 3m long time of flight tube for the transport of positron annihilation induced Auger electrons to the electron detector. The electron and positron flight paths are separated using an EXB separator. The annihilation gamma are detected using a HpGe detector and NaI(Tl) scintillation detector. The data shown here were collected by an HpGe detector in coincidence with the NaI(Tl) detector.

## RESULTS AND DISCUSSION

Figure 2 shows a comparison between the Doppler broadened annihilation gamma spectra obtained at positron energy of 20 keV and at positron energy of 2eV from MLG on Cu. It is clearly seen that there is a large difference between the Doppler broadened gamma spectra obtained at the two energies. Based on previous studies, [9] of the diffusion of positrons in well-annealed Cu after implantation using a variable energy beam, we believe that the data obtained at 20keV represent the Doppler broadened spectra from the bulk of the sample (in this case Cu). Our previous investigations of the intensities of the Auger peak in the positron annihilation induced Auger electron spectra (PAES) obtained with low energy positrons ( ~ 2 eV) [8, 10] indicate that the data obtained at 2 eV represent the Doppler broadened annihilation spectra solely from the surface (in this case MLG). It may be seen from Fig. 2 that the spectrum taken at 2 eV is much narrower than the spectra obtained at 20 keV. When the positron is in the surface state, there is a greater probability for the positron wave function to overlap with the valence electron wave function that extends out into the vacuum. Reduced core annihilation probability and increased valence annihilation can result in narrower gamma spectra. Positrons in the surface state do annihilate with core electrons, which forms the principle of PAES, and such annihilations would leave a signature through the corresponding broadening of the annihilation gamma spectra. In a clean Cu surface, even when the positrons are in the surface state there is some probability (though lesser than in bulk) for positrons to annihilate with 3d electrons, producing 3d electron momentum induced broadening of the gamma spectra. However, if the positron annihilates with only electrons of the carbon over layers, then the gamma spectra will reflect a complete absence of 3d electron momentum information. Hence, by looking at the amount of reduction in the broadening of the annihilation gamma it may be possible to differentiate between a surface annihilation induced narrowing and a narrowing due to changes in the top layer chemical structure. In order to qualitatively look at these differences at the high energy region of the 511 keV gamma spectra, the ratio curve method proposed in Asok-Kumar et al. [3] has been used and is discussed below.

The ratio of the gamma spectra taken at 2 eV to the gamma spectra at 20 keV is shown in Fig. 3. Only the high energy region of the 511 keV gamma spectra has been used to produce the ratio curve. The ratio curve quantitatively expresses the change in shape of the gamma spectra with the positron implantation energy. The experimental ratio curve is compared to a ratio between the theoretically calculated Doppler broadening spectrum from the bulk of graphitic carbon to the theoretically calculated gamma spectra from the bulk of Cu. Details of the calculation have been discussed in detail in previous publications [11, 12]. The theoretical gamma spectra have been convoluted with a Gaussian representing the energy resolution of the detector at 511 keV. The theoretically generated ratio curve compares very well with the experimentally generated ratio curves from pure bulk samples reported by Brusa et al. [4].

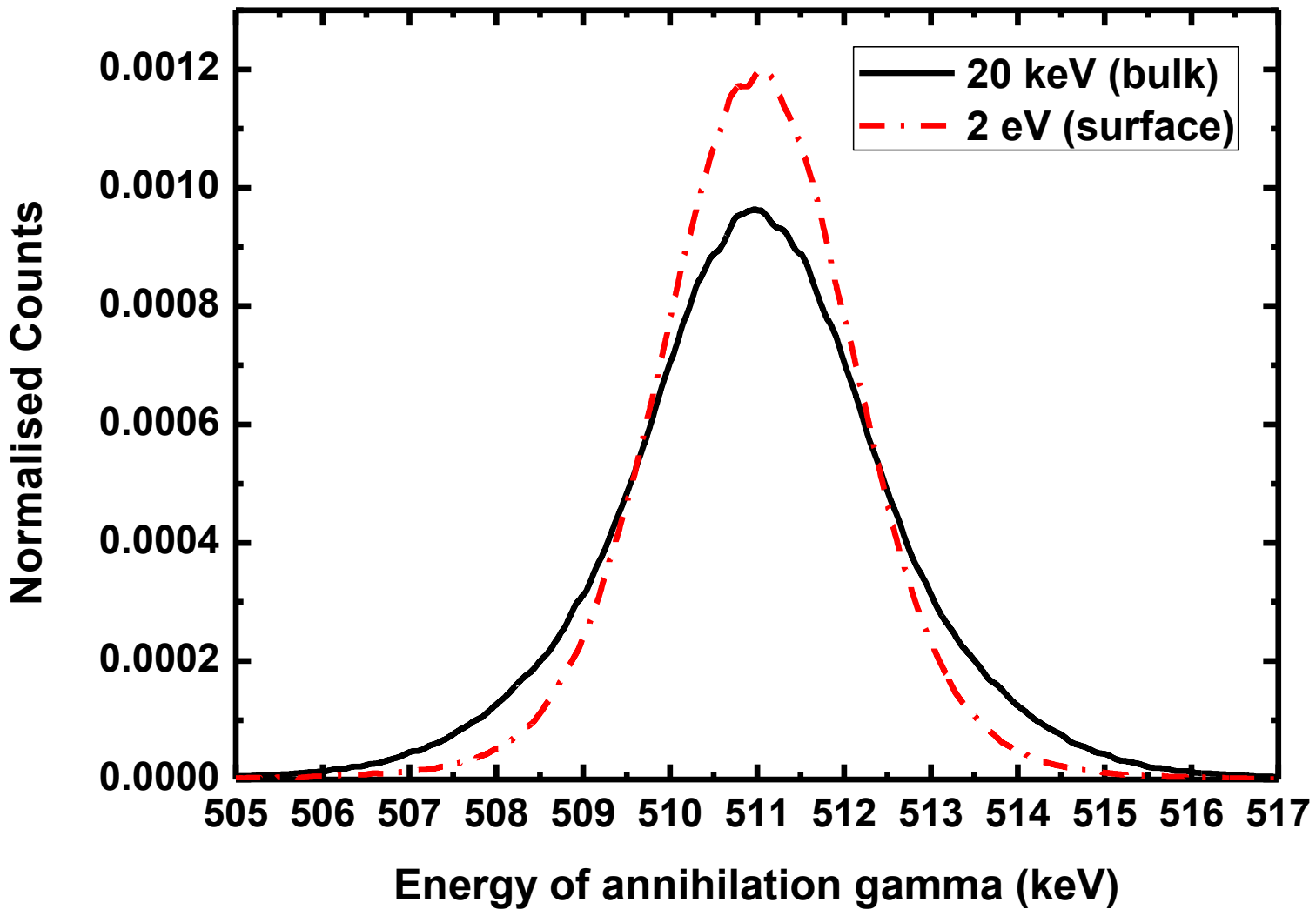

**FIGURE 2.** The 511 keV annihilation gamma spectrum collected from multi-layer graphene (MLG) on Cu at two positron energies (at 2 eV shown as dotted red line and at 20 keV shown as solid black line). Positrons implanted at 20 keV are annihilating in Cu (bulk) and positrons implanted at 2 eV are annihilating at the surface, which in this case is 6-8 over layers of graphene on Cu. A background has been subtracted from both spectra using the algorithm given by Mogensen et al. [13]. The difference in the two gamma spectra is consistent with the fact that positrons annihilate predominantly with valence electrons extending out into the vacuum when in the surface. However, the high energy region of the gamma spectrum from positrons annihilating at the surface will contain information on the chemical structure of the top surface.

The ratio between the gamma spectrum collected at 2 eV to the gamma spectrum collected at 20 keV differs from the theoretically generated ratio curve at the low momentum region (less than $10 \times 10^{-3} m_0 c$ ). This difference in the low momentum region is expected to be due to the difference in the electron cloud sampled by the positron in the bulk of a graphite sample and on the surface of graphene. The difference in the low momentum region can also be due to the contribution from para Ps annihilation to the experimental gamma spectra obtained at 2 eV. The gamma spectrum taken at 20 keV may also have (less than 3%) contributions from surface annihilations. However, the experimental ratio curve has an overall agreement to the dip seen in the theoretical ratio curve at ~ $15 \times 10^{-3} m_0 c$. The dip corresponds to the complete absence of 3d electron annihilations in the theoretical Carbon spectrum [4]. The overall agreement at higher momentum shows that the changes in the high energy region of the gamma spectra collected with positrons implanted at 2 eV corresponds to changes in chemical structure at the surface, which in this case is the presence of a few layers atomic layers of carbon. Since the 511 keV annihilation gamma can easily penetrate samples and the vacuum chamber walls that are a few tens of millimeter thick without any loss of resolution or electron momentum information, a careful analysis of the high energy region of the Doppler broadened annihilation gamma should be able to identify chemical changes on the first few atomic layers of external or more importantly internal surfaces in nanoporous materials that are used in catalysis, electrochemical or photo electrochemical reactions. The information on the chemical changes in the inner surfaces of nanoporous materials is presently inaccessible with conventional experimental techniques and beam based Doppler broadening spectroscopy attached to positron beams with the ability to carry out in-operando measurements can play an important role in the area of catalytic chemistry.

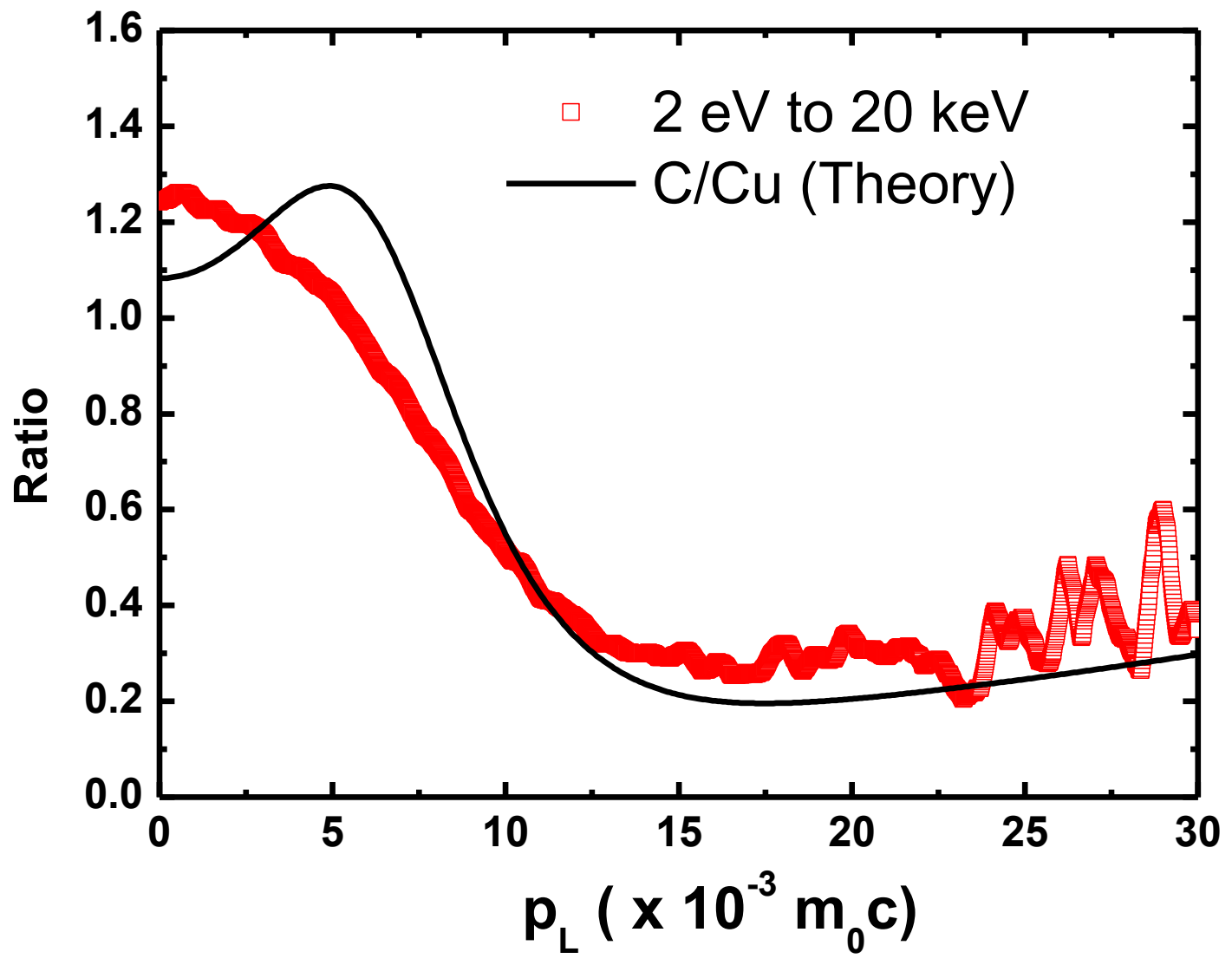

**FIGURE 3.** A ratio curve in which we have divided the Doppler broadened gamma spectra from the surface of Cu with 6-8 overlayers of Graphene (2 eV) by the Doppler broadened spectrum of bulk Cu (20 keV) (red squares). The solid black line is the ratio curve obtained by dividing the theoretically generated Doppler broadened spectrum from bulk graphitic carbon by the theoretically generated Doppler broadened spectrum bulk Cu. The overall agreement in the high momentum region shows that surface annihilation will reflect the chemical structure at the top most atomic layer.

## CONCLUSIONS

We have presented results from the investigations of the surface of a polycrystalline Cu with 6-8 layers of graphene using Doppler broadening spectroscopy. The spectra of Doppler broadened annihilation gamma rays were obtained at beam energies ranging from 2 eV up to 20 keV. The spectra obtained at 2 eV were observed to be much narrower in their momentum distribution than the spectra obtained at 20 keV. The narrowing observed in the 2eV data as compared to the 20keV spectra can be accounted for by positing that the positrons implanted at 2 eV become trapped in a surface state and annihilate with C atoms of the graphene overlayer, whereas the positrons incident at 20 keV are implanted deep enough into the bulk such that the majority do not live long enough to diffuse back to the surface and annihilate with the Cu substrate. We compared the ratio curves formed by dividing the measured Doppler broadened gamma spectra obtained at low positron kinetic energies (~2 eV) to the gamma spectra obtained at 20 keV to ratio curves found by dividing the calculated spectra of bulk graphite to bulk Cu. The ratio curves obtained from the measured results show qualitative agreement with those obtained from the calculated spectra and, in particular, both sets of curves indicate a much reduced intensity at high momentum. The qualitative agreement between the measured and calculated ratio curves provides further evidence that the Doppler broadened gamma spectra obtained at 2 eV is characteristic of the graphene layer at the surface. Additional measurements of the Doppler broadened gamma spectra are planned for clean and adsorbate covered external surfaces. The preliminary measurements shown in this paper provide strong evidence that it will be possible to obtain chemically sensitive information from the top atomic layers of surfaces (both internal and external) from an analysis of the high momentum tail of the Doppler broadened gamma spectra obtained from the annihilation of positrons bound at the surface and point to the possibility of developing an in-situ and non-destructive method of surface analysis suitable for investigating the chemical properties of the internal surfaces of porous materials in-operando conditions. This method will provide unique information not obtainable by other means because of its ability to selectively probe the chemically active top layers of internal surfaces of porous materials with minimal contributions to the spectroscopic signals from the non-chemically active bulk materials.

## ACKNOWLEDGMENTS

This work was supported by NSF grants DMR 1508719 and DMR 1338130, and Welch Foundation grant No. Y-1968-20180324.